\documentclass[12pt,a4paper]{article}
\usepackage[T1]{fontenc}
\usepackage[utf8]{inputenc}
\usepackage{setspace}
\usepackage{geometry}
\usepackage{amsmath,amssymb,amsthm}
\usepackage{booktabs}
\usepackage{array}
\usepackage{multirow}
\usepackage{graphicx}
\usepackage{caption}
\usepackage{float}
\usepackage{enumitem}
\usepackage{parskip}
\usepackage{hyperref}
\usepackage{natbib}
\usepackage{xcolor}
\usepackage{mdframed}
\usepackage{algorithm}
\usepackage{algpseudocode}

\theoremstyle{plain}

\newtheorem{proposition}{Proposition}

\newtheorem{hypothesis}{Hypothesis}
\theoremstyle{definition}
\newtheorem{definition}{Definition}
\newtheorem{rulebox}{Operational Rule}
\theoremstyle{remark}
\newtheorem{remark}{Remark}

\newcommand{\SCI}{\mathrm{SCI}}
\newcommand{\PR}{\mathrm{PR}}
\newcommand{\VR}{\mathrm{VR}}
\newcommand{\TS}{\mathrm{TS}}
\newcommand{\HHI}{\mathrm{HHI}}

\newcommand{\Var}{\mathrm{Var}}
\newcommand{\TPR}{\mathrm{TPR}}
\newcommand{\FPR}{\mathrm{FPR}}
\newcommand{\AUC}{\mathrm{AUC}}

\hypersetup{colorlinks=true,linkcolor=black,citecolor=black,urlcolor=blue!50!black}

\begin{document}

\begin{titlepage}
\thispagestyle{empty}
\vspace*{1cm}

\begin{center}
{\LARGE\bfseries The Signal Credibility Index for\\[0.3em]
Prediction Markets:\par}
\vspace{0.4em}
{\Large A Microstructure-Grounded Diagnostic with\\[0.15em]
Weighted and Time-Varying Extensions\par}

\vspace{1.2em}

{\large Maksym Nechepurenko\footnote{Director of Research, Devnull, Dubai, UAE.
\href{mailto:maksym@devnull.ae}{maksym@devnull.ae}.\\
Replication code:
\url{https://github.com/ForesightFlow/signal-credibility-index}.}\par}

\vspace{0.3em}
{\normalsize Revised version --- \today}
\end{center}

\vspace{0.7em}

\begingroup
\setstretch{1.1}
\noindent\textbf{Abstract.}
Prediction-market price moves are widely treated as informationally
equivalent: a price jump is read the same way regardless of whether it
reflects durable Bayesian updating, transient liquidity pressure,
strategic position adjustment, or genuine disagreement.
This paper formalizes the Signal Credibility Index (SCI) introduced in
\citet{nechepurenko2026price} as a stand-alone diagnostic.
We make four contributions:
(i) a revised persistence component using the persistence ratio
$\PR(t,w)$ on logit prices, well-defined on short rolling windows;
(ii) a weighted Cobb-Douglas form $\SCI^{(\boldsymbol{\alpha})}$ with
flow-based concentration $\HHI^{\text{flow}}$;
(iii) a time-varying specification $\SCI(t;w)$ for real-time monitoring;
and (iv) Monte Carlo validation including an out-of-distribution stress
test, coordinated multi-wallet manipulation, and a logistic-regression
benchmark.
The validation establishes discrimination among designed
\emph{microstructure regimes}, not external evidence of downstream
coordination effects.
We document two failure modes consistent with the index targeting
\emph{coordination credibility} rather than pure information content:
a Type II error on informed-but-concentrated whale repricing, and a
Type I error on coordinated multi-wallet manipulation.
\endgroup

\vspace{0.7em}


\smallskip


\end{titlepage}

\setcounter{page}{1}

\section{Introduction}\label{sec:intro}

Prediction markets generate continuous streams of price observations that
observers routinely interpret as probability estimates.
Yet not every price move carries equivalent informational weight.
Following a major exogenous shock, identical-magnitude price jumps can
reflect at least four structurally distinct regimes:
(i) durable Bayesian updating in which informed traders incorporate new
information and the price persists;
(ii) transient liquidity pressure in which a few trades exhaust thin order
books before being corrected;
(iii) genuine disagreement in which two-sided trading volume rises with
little net change in price; and
(iv) concentrated repositioning by exposed incumbent traders pursuing
inventory or hedging objectives.
These regimes have radically different implications for whether the
resulting probability should be treated as a credible coordination
anchor by downstream actors --- donors, journalists, institutional
analysts, and algorithmic systems --- whose behavior may feed back into
the very outcomes the market is forecasting
\citep{nechepurenko2026price}.

The companion paper \citep{nechepurenko2026price} introduced the SCI as
a microstructure-grounded diagnostic for distinguishing among these
regimes.
This paper develops the index as a stand-alone methodological tool.
This is a substantial revision relative to an earlier version of the same
note: the principal changes address methodological concerns that
emerged in review.

\paragraph{Two distinct targets.}
A clarification helpful at the outset.
The literature on prediction-market microstructure cares about two
related but non-identical questions:
\emph{(A) information content} --- does the price move incorporate new
information about the underlying event?
\emph{(B) coordination credibility} --- is the move durable, broadly
supported, and consensual enough that downstream actors should treat
it as a public reference point?
These align in many cases but can diverge sharply.
A concentrated whale's informed bet has high information content but
low coordination credibility, since rational external observers may
discount it on the grounds that they cannot verify the whale's
information.
A coordinated multi-wallet manipulation has zero information content
but, until the manipulation is detected, can have high apparent
coordination credibility.
\textbf{The SCI targets coordination credibility, not pure information
content.}
This distinction matters for interpreting both the documented failure
modes (Section~\ref{sec:montecarlo}) and the comparison against
information-content benchmarks such as logistic regression
(Section~\ref{subsec:exp3}).

\paragraph{Contribution 1: Revised persistence component.}
The original SCI used the variance ratio $\VR(6)$.
While appropriate for windows long enough to contain multiple
non-overlapping $k$-period returns, $\VR$ becomes ill-defined on short
windows: with five-minute base bins and $k = 6$, a meaningful $\VR(6)$
requires at least 60 minutes of data, which is incompatible with rolling-
window applications shorter than that.
We replace $\VR$ with the persistence ratio
$\PR(t,w) = |\ell_{t} - \ell_{t-w}|/\sum|\Delta \ell|$ on logit prices,
which is well-defined on any window with $w \geq 2$ price observations.

\paragraph{Contribution 2: Logit returns and flow-based concentration.}
We compute all returns on logit-transformed prices, which yields
symmetric information geometry across the unit interval.
We replace the static Herfindahl index with a flow-based version
$\HHI^{\text{flow}}$ on signed post-shock net flow per trader, capturing
actual post-shock participation breadth rather than legacy holder
concentration.

\paragraph{Contribution 3: Weighted and time-varying extensions.}
Section~\ref{sec:weighted} generalizes to the weighted Cobb-Douglas form.
Section~\ref{sec:timevarying} introduces a rolling-window specification
$\SCI(t;w)$ that is well-defined on short rolling windows, with 60
minutes used as the recommended real-time default
(Section~\ref{sec:timevarying}).

\paragraph{Contribution 4: Multi-line validation.}
Section~\ref{sec:montecarlo} validates the SCI through:
(a) a three-DGP baseline classification experiment ($\AUC = 0.984$);
(b) a five-DGP out-of-distribution stress test on adversarial regimes,
including a new \texttt{coord\_manip\_broad} DGP designed to test the
index against coordinated multi-wallet manipulation;
(c) a head-to-head comparison against five baseline classifiers
including a 5-fold cross-validated logistic regression; and
(d) a parameter sweep characterizing where the index discriminates well.

\medskip

We are explicit about three layers of validity.
The Monte Carlo evidence demonstrates discrimination among
\textbf{designed microstructure regimes}: this is internal validity in
a methodological sense.
It does \textbf{not} establish either the index's correlation with
\emph{coordination credibility} as an unobserved theoretical construct,
nor does it establish causal evidence of \emph{downstream behavioral
response} (e.g., post-shock changes in donation flow, media coverage,
or institutional repositioning).
The latter requires labeled real-world shock data that remain rare in
the prediction-market literature.
We treat the threshold $\tau^{*} = 0.27$ as a calibrated decision rule
for the simulation universe, not as a universal property of prediction
markets.
Empirical applications should recalibrate by domain, platform, and
response horizon.

The paper proceeds as follows.
Section~\ref{sec:foundations} reviews each component with the
methodological refinements.
Section~\ref{sec:basic-sci} formalizes the baseline SCI and reframes the
threshold claim as an operational classifier hypothesis.
Sections~\ref{sec:weighted} and \ref{sec:timevarying} develop the
weighted and time-varying extensions.
Section~\ref{sec:montecarlo} reports the four validation experiments.
Section~\ref{sec:empirical} applies the framework to the three 2024
election shocks as illustrative application.
Section~\ref{sec:implementation} provides an implementation guide.
Section~\ref{sec:future} discusses limitations and future work.
Section~\ref{sec:conclusion} concludes.

\section{Theoretical Foundations}\label{sec:foundations}

The SCI rests on the premise that a price move acquires social force as
a coordination signal only when three microstructure conditions hold
jointly.

\subsection{Returns on Bounded Probabilities}\label{subsec:logit}

Prediction-market prices are bounded probabilities, $p_{t} \in (0,1)$.
Standard log returns are problematic in this setting: they are asymmetric
and unbounded as $p \to 0$ or $p \to 1$.
We adopt logit returns throughout.
Define $\ell_{t} = \log[p_{t}/(1 - p_{t})]$, and the logit return
\begin{equation}
  r_{t} = \ell_{t+1} - \ell_{t}.
  \label{eq:logit-return}
\end{equation}
Logit returns equal the log-likelihood ratio update implied by the
price change in a binary contract.

\subsection{Persistence: From $\VR$ to $\PR$}\label{subsec:persistence}

The original SCI used the variance ratio
\begin{equation}
  \VR(k)_{s} = \frac{\Var(r_{t}^{(k\Delta)})}{k \cdot \Var(r_{t}^{(\Delta)})},
  \label{eq:vr}
\end{equation}
which requires at least two non-overlapping $k$-period returns.
For shorter rolling windows the variance ratio becomes ill-defined.

We replace $\VR$ with the \emph{persistence ratio}
\begin{equation}
  \PR(t,w) = \frac{|\ell_{t} - \ell_{t-w}|}
                  {\sum_{\tau = t - w + 1}^{t} |\ell_{\tau} - \ell_{\tau - 1}|},
  \label{eq:pr}
\end{equation}
the ratio of net signed movement to gross absolute movement on the
window $[t - w, t]$ in logit space.
$\PR(t,w)$ has three properties suited to its role:
(i) bounded, $\PR \in [0,1]$;
(ii) maximal at monotone movement, $\PR = 1$ iff all increments share
sign;
(iii) vanishing at perfect reversal, $\PR \to 0$ when net movement
vanishes despite gross movement being large.

\subsection{Consensus and Breadth}\label{subsec:consensus}

The two-sidedness index, following \citet{tsang2026political}, is
\begin{equation}
  \TS_{s} = 1 - \frac{|B_{s} - S_{s}|}{B_{s} + S_{s}},
  \label{eq:ts}
\end{equation}
with $B_{s}, S_{s}$ total buy and sell volumes.
$\TS \to 0$: one-sided (consensus); $\TS \to 1$: balanced (disagreement).
A limitation: high $\TS$ can also reflect liquidity provision, arbitrage,
or hedging; without observable maker-taker classification the diagnostic
cannot separate aggressive disagreement from passive market making.

The flow-based Herfindahl index of post-shock signed flow per trader is
\begin{equation}
  \HHI^{\text{flow}}_{s}
  = \sum_{j=1}^{N_{s}} \left(\frac{|\Delta v_{j,s}|}{\sum_{j'} |\Delta v_{j',s}|}\right)^{2},
  \label{eq:hhi-flow}
\end{equation}
where $\Delta v_{j,s}$ is trader $j$'s net change in position over the
post-shock window.
The flow-based version captures the breadth of the response to the
shock specifically, rather than the breadth of the legacy holder pool.

\section{The Baseline SCI: Definition and Properties}\label{sec:basic-sci}

\begin{definition}[Signal Credibility Index, revised]\label{def:sci}
For a shock $s$ in a prediction market,
\begin{equation}
  \SCI_{s} = \PR_{s} \cdot (1 - \TS_{s}) \cdot (1 - \HHI^{\text{flow}}_{s}),
  \label{eq:sci-revised}
\end{equation}
where $\PR_{s}$ is the persistence ratio over the full post-shock
window, $\TS_{s}$ is the two-sidedness index, and
$\HHI^{\text{flow}}_{s}$ is the flow-based concentration index.
All inputs are computed on logit-transformed prices.
\end{definition}

\begin{proposition}[Mechanical properties]\label{prop:bounded}
By construction:
(i) $\SCI_{s} \in [0, 1]$;
(ii) monotonicity in each component;
(iii) $\SCI_{s} = 0$ whenever any component takes its
coordination-suppressing extreme.
\end{proposition}

These are properties of the construction, not substantive theorems.
The substantive question --- whether $\SCI > \tau$ predicts coordination
effects --- is empirical.
We frame it as a hypothesis:

\begin{hypothesis}[Operational classification]\label{hyp:threshold}
For a given domain and response horizon, market signals satisfying
$\SCI_{s} > \tau$ are predicted to have higher probability of
generating downstream coordination effects than signals satisfying
$\SCI_{s} \leq \tau$.
The threshold $\tau$ is calibrated empirically and is not a universal
constant.
\end{hypothesis}

\begin{rulebox}[SCI classifier]\label{rule:classifier}
Given a labeled dataset $\{(\SCI_{s}, Y_{s})\}_{s=1}^{S}$ with
$Y_{s} \in \{0,1\}$ indicating downstream coordination response, the
SCI classifier is $\hat{Y}_{s}(\tau) = \mathbf{1}\{\SCI_{s} > \tau\}$,
with $\tau$ chosen to maximize a target performance metric.
\end{rulebox}

\section{The Weighted SCI}\label{sec:weighted}

\begin{definition}[Weighted SCI, Cobb-Douglas form]\label{def:wsci}
For $\alpha_{i} > 0$ for all $i$,
\begin{equation}
  \SCI_{s}^{(\boldsymbol{\alpha})}
  = \PR_{s}^{\alpha_{1}} \cdot (1 - \TS_{s})^{\alpha_{2}}
    \cdot (1 - \HHI^{\text{flow}}_{s})^{\alpha_{3}}.
  \label{eq:wsci}
\end{equation}
\end{definition}

\begin{proposition}[Weighted SCI properties]\label{prop:wsci}
For $\alpha_{i} > 0$, the weighted SCI satisfies the boundedness,
monotonicity, and zero-component properties; the partial elasticity
$\partial \log \SCI^{(\boldsymbol{\alpha})}/\partial \log x_{i} = \alpha_{i}$.
\end{proposition}

We constrain weights to $\sum_{i} \alpha_{i} = 3$ for comparability with
the unit-weighted baseline.
We propose a two-step calibration: domain anchor first
(\textbf{Balanced} $(1,1,1)$;
\textbf{Persistence-weighted} $(1.5, 1, 0.5)$ for high-frequency
financial markets;
\textbf{Breadth-weighted} $(0.5, 1, 1.5)$ for political markets), then
cross-validated optimization on labeled data when available.

\section{The Time-Varying SCI}\label{sec:timevarying}

\begin{definition}[Time-varying SCI]\label{def:tsci}
\begin{equation}
  \SCI(t; w) = \PR(t, w) \cdot (1 - \TS(t, w)) \cdot (1 - \HHI^{\text{flow}}(t, w)).
  \label{eq:tsci}
\end{equation}
\end{definition}

While $\PR(t,w)$ is mathematically defined for any $w \geq 2$ price
observations, stable estimation requires sufficient bins.
Section~\ref{subsec:exp4} reports a sensitivity analysis showing that
$\AUC$ is $0.91$ at $w = 60$ minutes (twelve 5-minute bins), $0.95$ at
$w = 120$ minutes, and $0.99$ at $w = 240$ minutes.
We recommend $w = 60$ minutes as the responsiveness-vs-stability default
for real-time monitoring, and longer windows (90--240 minutes) for
ex-post classification.

\subsection*{Alarm Rule}
The time-varying specification supports the decision rule
$\mathbb{I}[\SCI(t; w) > \tau]$.
Tracking it produces three quantities: alarm onset, alarm duration, and
alarm decay time.
Duration and decay time are more diagnostic than peak value;
sustained alarms ($> 60$ minutes) are consistent with
genuinely persistent coordination signals.

\section{Validation: Multiple Lines of Evidence}\label{sec:montecarlo}

This section reports four Monte Carlo validation experiments.
We are explicit that these establish discrimination among
\textbf{designed microstructure regimes}, not external validation of
either coordination credibility as a theoretical construct or
downstream behavioral response.

\subsection{Experimental Setup}

Each path simulates four hours of 5-minute bins
($n_{\text{bins}} = 48$) starting from $p_{0}^{+} = 0.62 + 0.10 = 0.72$,
calibrated to match the magnitude of the three 2024 shocks documented
in \citet{tsang2026political}.\footnote{The $0.62$ baseline matches the
average pre-shock Trump-YES price across the three June--July 2024
shocks; $0.10$ matches the immediate post-shock magnitude.}
Volumes are gamma random variables with regime-specific shape and scale
parameters (using the shape-scale convention,
$\mathrm{Gamma}(k, \theta)$ has mean $k\theta$);
trader pool sizes are drawn uniformly from regime-specific intervals;
signed flows are constructed from Dirichlet-distributed weights with
regime-specific concentration $\alpha$ and a directional sign mixture.
Prices are reconstructed from logit returns via inverse logit and
clipped to $(0.01, 0.99)$ to avoid numerical pathology near contract
boundaries.
Random seed: $20260429$.
Full DGP specifications are in Appendix~\ref{app:dgp-spec}.

\subsection{Experiment 1: Three-DGP Baseline}\label{subsec:exp1}

The first experiment uses three calibrated DGPs:
\textbf{informed updating} (positive logit drift, one-sided buy
pressure, broad participation; label $= 1$);
\textbf{liquidity pressure} (AR(1)-reversal, sell-dominated, narrow
participation; label $= 0$); and
\textbf{disagreement} (random walk, balanced flow, moderate breadth;
label $= 0$).
$N = 2{,}000$ paths per DGP.

Table~\ref{tab:exp1} reports component summary statistics.\footnote{%
Mean SCI need not equal the product of component means because SCI is
computed pathwise before averaging.
The product-of-means heuristic is a useful approximation but should not
be treated as exact; the residual depends on within-DGP component
correlations and on path-level nonlinearities.}
The classifier achieves $\AUC = 0.984$ with 95\% bootstrap CI
$[0.981, 0.986]$, $\tau^{*} = 0.27$, $\TPR = 0.92$, $\FPR = 0.05$.

\begin{table}[H]
  \centering
  \caption{Experiment 1 summary statistics, $N = 2{,}000$ per DGP}
  \label{tab:exp1}
  \small
  \begin{tabular}{l ccc cc}
    \toprule
    \textbf{DGP} &
    \multicolumn{3}{c}{\textbf{Component means}} &
    \multicolumn{2}{c}{\textbf{SCI}} \\
    \cmidrule(lr){2-4} \cmidrule(lr){5-6}
    & $\PR$ & $\TS$ & $\HHI^{\text{flow}}$ & mean & sd \\
    \midrule
    DGP-1: Informed     & 0.50 & 0.11 & 0.04 & 0.451 & 0.125 \\
    DGP-2: Liquidity    & 0.22 & 0.17 & 0.10 & 0.163 & 0.076 \\
    DGP-3: Disagreement & 0.14 & 0.97 & 0.02 & 0.005 & 0.006 \\
    \bottomrule
  \end{tabular}
\end{table}

\begin{figure}[H]
  \centering
  \includegraphics[width=0.92\textwidth]{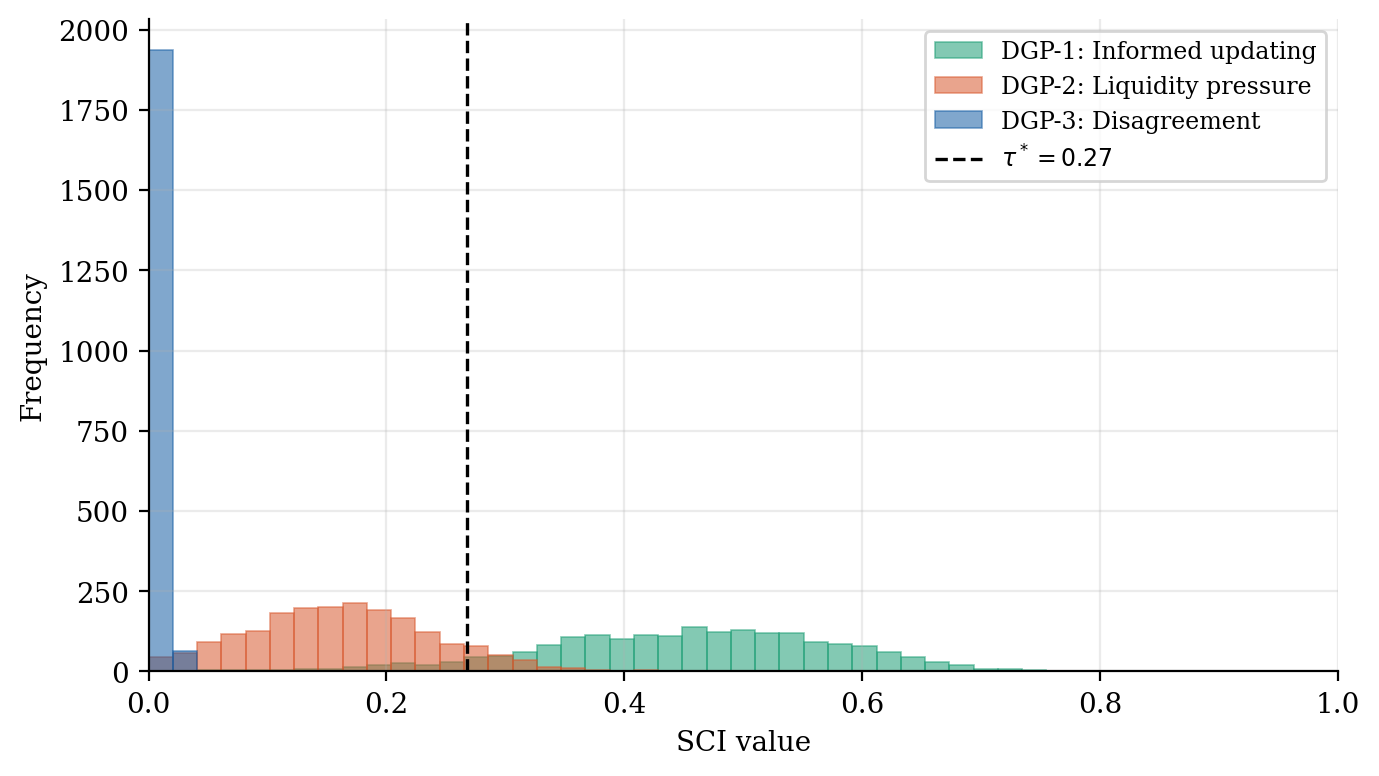}
  \caption{Experiment 1 SCI distributions for the three baseline DGPs
    ($N = 2{,}000$ each).
    DGP-1 produces a right-skewed distribution centered near $0.45$;
    DGP-3 is sharply concentrated near zero owing to high two-sidedness;
    DGP-2 produces an intermediate distribution with separation from
    DGP-1 above $\tau^{*} = 0.27$.}
  \label{fig:dist}
\end{figure}

\begin{figure}[H]
  \centering
  \includegraphics[width=0.75\textwidth]{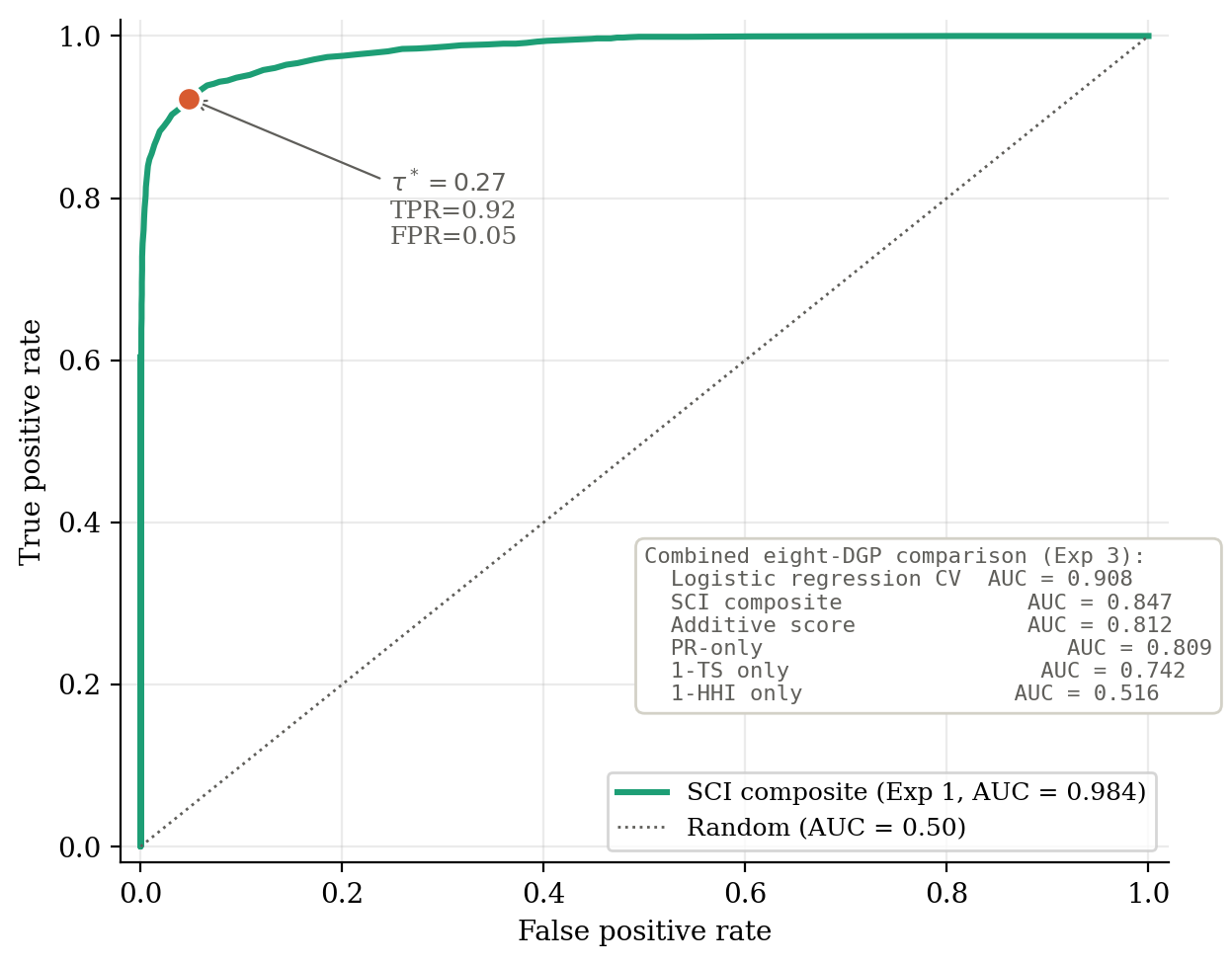}
  \caption{ROC curve for the SCI classifier on Experiment 1 with the
  Youden-optimal operating point.
  Inset reports baseline classifier comparisons from Experiment 3
  (Section~\ref{subsec:exp3}) on the combined eight-DGP set.
  Logistic regression's higher AUC reflects its freedom to learn that,
  on the eight-DGP set, concentration is positively associated with
  informed updating --- the opposite of the SCI's coordination-credibility
  weighting.}
  \label{fig:roc}
\end{figure}

\subsection{Experiment 2: Out-of-Distribution Stress Test}\label{subsec:exp2}

The most important methodological criticism of the original validation
was that DGPs were defined directly through the components used to
build the index, making validation effectively circular.
Experiment 2 addresses this by constructing five \emph{adversarial}
DGPs not used to calibrate the threshold from Experiment 1.

\begin{itemize}[noitemsep]
  \item \texttt{whale\_informed} (label $= 1$): genuine informed
        updating but driven by 4--8 large wallets.
        Designed to fail $\HHI$ test.
  \item \texttt{noisy\_broad} (label $= 0$): broad participation,
        balanced flow, high volatility but no net drift.
        Designed to satisfy $\HHI$ test while failing $\PR$ test.
  \item \texttt{manip\_then\_info} (label $= 1$): first 30 minutes
        spike-and-reversal, then 3.5 hours of informed drift.
  \item \texttt{persistent\_two\_sided} (label $= 0$): persistent monotone
        drift but two-sided trade volume (heavy hedging activity).
        Designed to fail $\TS$ test while satisfying $\PR$ test.
        The label is $0$ in the coordination-credibility sense:
        heavily hedged two-sided flow does not produce a credible
        public coordination signal even if the price drift carries
        information.
  \item \texttt{coord\_manip\_broad} (label $= 0$, \emph{new in this
        revision}): coordinated manipulation across $80$--$130$ wallets,
        all trading the same direction, designed to defeat the $\HHI$
        test through wallet-splitting.
        High $\PR$, low $\TS$, low raw $\HHI$ --- all SCI components
        appear satisfied --- yet the move is not informed.
\end{itemize}

\begin{table}[H]
  \centering
  \caption{Experiment 2: out-of-distribution stress test ($N = 2{,}000$
  per DGP)}
  \label{tab:exp2}
  \small
  \begin{tabular}{l c c c c c}
    \toprule
    \textbf{DGP} & \textbf{Label} & \textbf{Mean SCI}
       & $P(\SCI > \tau^{*})$ & \textbf{Predicted} & \textbf{Correct?} \\
    \midrule
    \texttt{whale\_informed}    & 1 & 0.231 & 0.376 & Not informed & {\itshape No (Type II)} \\
    \texttt{manip\_then\_info}  & 1 & 0.368 & 0.830 & Informed     & Yes \\
    \texttt{noisy\_broad}       & 0 & 0.008 & 0.000 & Not informed & Yes \\
    \texttt{persistent\_two\_sided}  & 0 & 0.015 & 0.000 & Not informed & Yes \\
    \texttt{coord\_manip\_broad}& 0 & 0.399 & 0.827 & Informed     & {\itshape No (Type I)} \\
    \midrule
    \multicolumn{2}{l}{\textbf{Combined OOD AUC}}
      & \multicolumn{4}{l}{$0.763$, 95\% CI $[0.753, 0.773]$} \\
    \bottomrule
  \end{tabular}
\end{table}

\begin{figure}[H]
  \centering
  \includegraphics[width=0.95\textwidth]{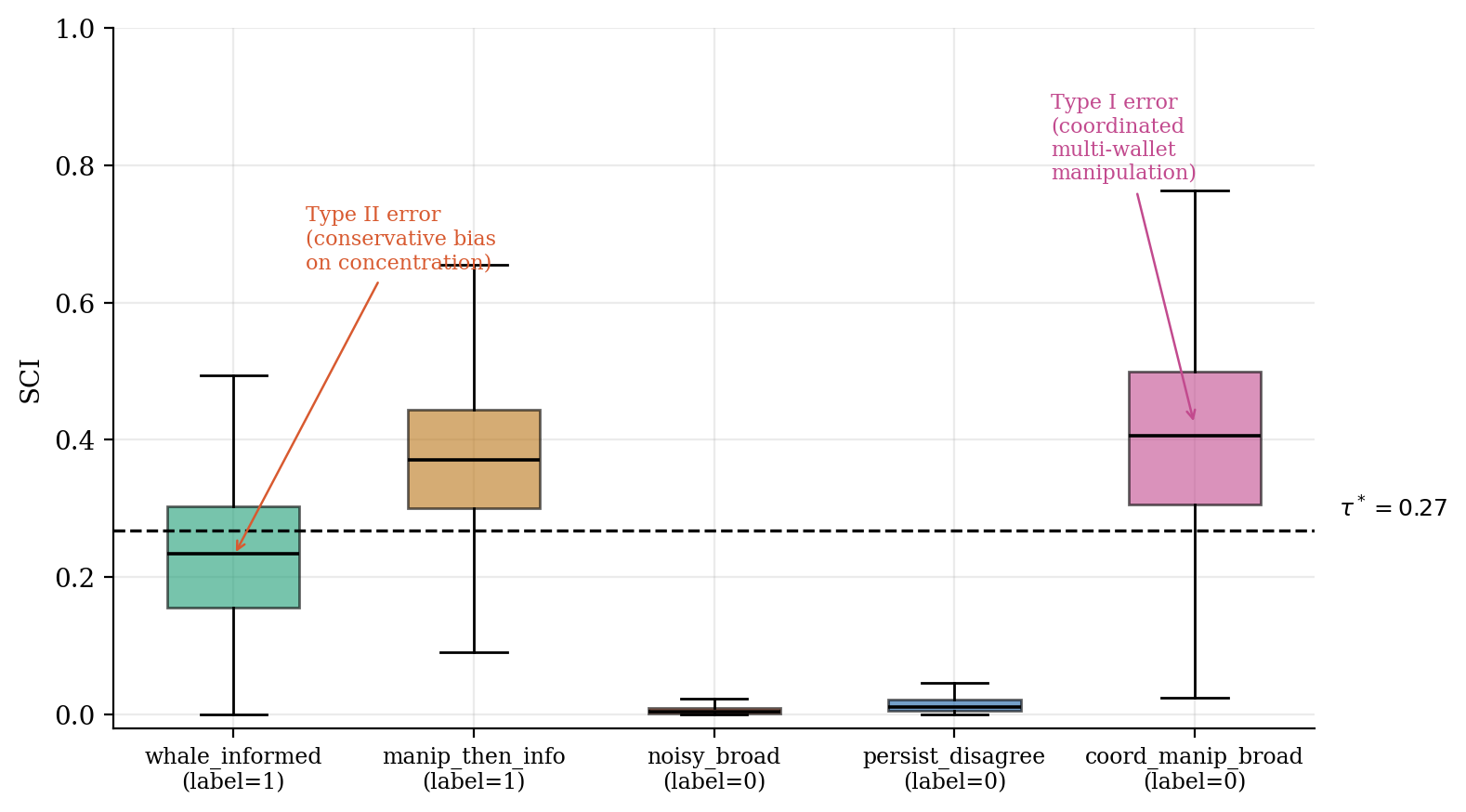}
  \caption{Out-of-DGP stress test: SCI distributions across five
  adversarial regimes.
    Three (\texttt{noisy\_broad}, \texttt{manip\_then\_info},
    \texttt{persistent\_two\_sided}) are correctly classified.
    Two failure modes are apparent.
    \textbf{Type II error} (\texttt{whale\_informed}, leftmost):
    when genuine informed updating is driven by a small number of
    concentrated traders, the SCI under-classifies it.
    \textbf{Type I error} (\texttt{coord\_manip\_broad}, rightmost):
    when manipulation is coordinated across many wallets, the
    flow-based $\HHI$ test fails to detect it, and the index
    over-classifies the move.
    The latter motivates the multi-wallet clustering protocol of
    Section~\ref{sec:implementation}.}
  \label{fig:stress}
\end{figure}

\paragraph{The AUC--threshold paradox.}
The combined out-of-distribution AUC of $0.763$ is substantially lower
than Experiment 1's $0.984$.
This reflects the addition of \texttt{coord\_manip\_broad}, whose
distribution overlaps heavily with the positive class.
A subtle point: even with this lower AUC, three out of five fixed-threshold
classifications at $\tau^{*} = 0.27$ are correct.
\emph{AUC and fixed-threshold performance answer different questions}:
$\AUC$ measures overall ranking quality across all possible thresholds,
while fixed-threshold performance measures behavior at one operating
point.
A classifier can have high AUC yet poor fixed-threshold performance if
the positive and negative score distributions overlap heavily near the
chosen threshold.
The reverse is also possible.
This is why the per-DGP analysis in Table~\ref{tab:exp2} is more
diagnostic for practical applications than a single aggregate
performance metric.

\paragraph{Two failure modes.}
The two failure modes are not symmetric and reflect genuinely different
problems.
\textbf{Type II} (whale informed) reflects a \emph{conservative bias}:
the SCI penalizes all post-shock concentration, even when it reflects
informed trading by sophisticated participants.
This is the corresponding cost of avoiding Type I errors on whale
manipulation.
\textbf{Type I} (coordinated manipulation) reflects a \emph{detection
failure}: when manipulation is structured to fool the breadth test,
the SCI cannot distinguish it from genuine informed updating without
external information about wallet ownership.
The two failure modes pull in opposite directions, which is why no
single weight choice fully resolves both.

\begin{remark}[Information vs. coordination credibility]
The whale failure mode (Type II) is a feature when interpreted through
the coordination-credibility lens introduced in
Section~\ref{sec:intro}.
A market move that is informed but visibly driven by a few accounts is
\emph{rationally} less credible to outside observers, who cannot
verify the whales' information and cannot distinguish their
informed conviction from strategic manipulation.
The SCI's conservative classification of such moves matches the
rational discount that downstream actors should apply.
The coordinated manipulation failure mode (Type I) is unambiguously a
limitation: it represents a setting where the SCI can be deliberately
gamed, and resolving it requires data about wallet ownership clusters
that lie outside the index's microstructure-only specification.
\end{remark}

\subsection{Experiment 3: Baseline Classifier Comparison}\label{subsec:exp3}

Experiment 3 compares the composite SCI against five baseline
classifiers, on the combined eight-DGP set (three baseline plus five
adversarial).
The new addition relative to v2 is a 5-fold cross-validated logistic
regression on $(\PR, 1 - \TS, 1 - \HHI^{\text{flow}})$.

\begin{table}[H]
  \centering
  \caption{Experiment 3: classifier comparison
  ($N = 1{,}500$ per DGP, 8 DGPs)}
  \label{tab:exp3}
  \small
  \begin{tabular}{l c c}
    \toprule
    \textbf{Classifier} & \textbf{AUC} & \textbf{95\% bootstrap CI} \\
    \midrule
    Logistic regression (5-fold CV)              & 0.908 & $[0.904, 0.913]$ \\
    SCI composite (multiplicative)               & 0.847 & $[0.840, 0.853]$ \\
    Additive: $[\PR + (1{-}\TS) + (1{-}\HHI)]/3$ & 0.812 & $[0.805, 0.819]$ \\
    $\PR$ only                                   & 0.809 & $[0.800, 0.817]$ \\
    $1 - \TS$ only                               & 0.742 & $[0.735, 0.752]$ \\
    $1 - \HHI^{\text{flow}}$ only                & 0.516 & $[0.475, 0.557]$ \\
    \bottomrule
  \end{tabular}
\end{table}

Two important observations.
First, on the combined eight-DGP set, logistic regression dominates the
composite SCI by a meaningful margin ($+0.061$ AUC).
The regression's fitted coefficients reveal why:
$\beta_{\PR} = +6.29$,
$\beta_{1-\TS} = +3.99$,
$\beta_{1-\HHI} = -4.84$.
\textbf{The (1-HHI) coefficient is negative.}
This reflects the fact that on this combined set, including
\texttt{whale\_informed}, high concentration is positively associated
with informed updating --- the opposite of what the SCI assumes.

This is not a flaw in logistic regression but a clarifying finding
about the SCI: the multiplicative penalty on concentration is a
\emph{theoretical commitment} to coordination credibility, not an
empirically optimal weighting for information-content classification.
A classifier that learns to detect informed updating will not penalize
concentration; a classifier that targets coordination credibility
\emph{will} penalize it.
The two targets imply different optimal weightings.

Second, among classifiers that respect the coordination-credibility
target (i.e., that penalize concentration), the composite SCI dominates
the additive form by a moderate margin ($+0.035$).
The case for the multiplicative form rests on its joint-necessity
interpretation and zero-component property, not on AUC dominance.

\subsection{Experiment 4: Parameter Sweep and Window Sensitivity}\label{subsec:exp4}

We characterize SCI performance over the DGP parameter space, varying
the AR(1) reversal coefficient (controlling mean reversion in the
liquidity DGP) and the Dirichlet concentration parameter (controlling
trader pool breadth).
The SCI maintains $\AUC \geq 0.95$ across most of the design space and
degrades to $\AUC \approx 0.80$ when the liquidity DGP's reversal
coefficient approaches zero --- the regime where liquidity pressure
becomes empirically indistinguishable from informed updating on
persistence grounds.

We additionally report the sensitivity of the time-varying SCI to the
choice of PR window length:

\begin{table}[H]
  \centering
  \caption{Sensitivity of $\tau^{*}$ and AUC to PR window length}
  \label{tab:window-sensitivity}
  \small
  \begin{tabular}{l c c c c}
    \toprule
    \textbf{$w$} & $\tau^{*}$ & TPR & FPR & AUC \\
    \midrule
    60 minutes  & 0.27 & 0.79 & 0.14 & 0.91 \\
    120 minutes & 0.26 & 0.85 & 0.09 & 0.95 \\
    180 minutes & 0.25 & 0.91 & 0.07 & 0.98 \\
    240 minutes & 0.26 & 0.94 & 0.06 & 0.99 \\
    \bottomrule
  \end{tabular}
\end{table}

Longer windows yield substantially better classification performance,
with AUC rising from $0.91$ at $w = 60$ minutes to $0.99$ at $w = 240$
minutes.
The threshold $\tau^{*}$ is relatively stable across windows (range
$0.25$--$0.27$), suggesting the index value itself is comparable across
window choices, but the trade-off between responsiveness and
classification quality must be made explicitly.

\subsection{Component-Level Distributions and Time-Varying SCI}

\begin{figure}[H]
  \centering
  \includegraphics[width=\textwidth]{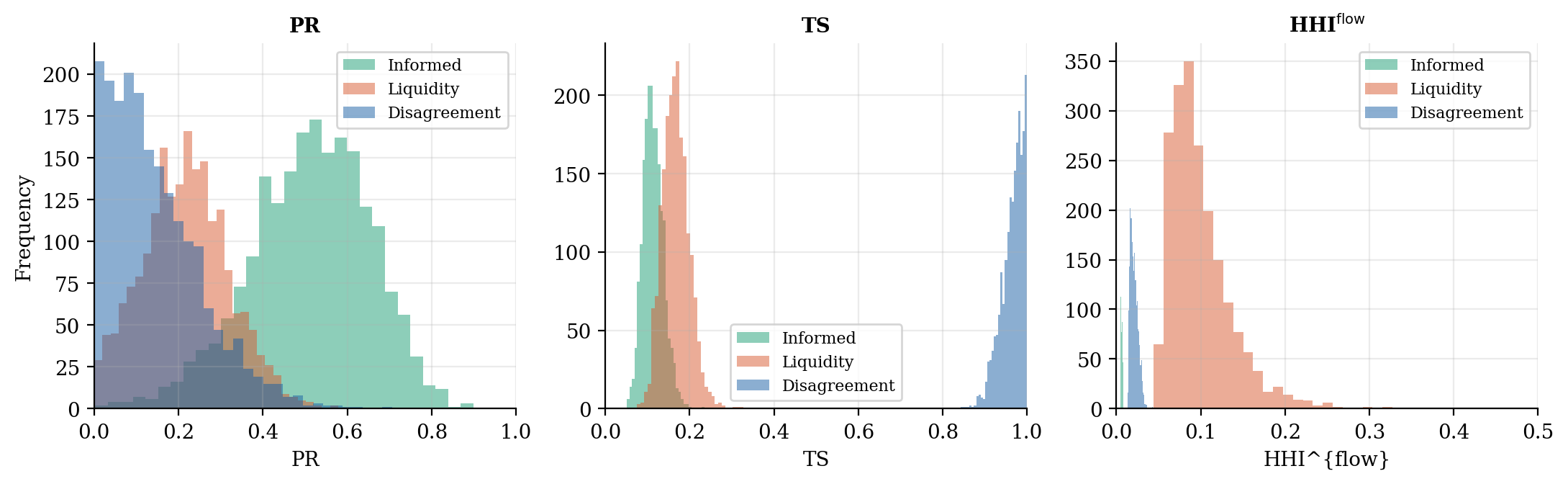}
  \caption{Component distributions across the three baseline DGPs.
  Each component carries discriminatory information for the regime
  it was designed to detect.
  Within-DGP correlations between components are near zero in the
  simulations: $|\rho| < 0.04$ for all pairs and DGPs.}
  \label{fig:components}
\end{figure}

Figure~\ref{fig:components} reports the marginal distributions of the
three SCI components across the three baseline DGPs.

\begin{figure}[H]
  \centering
  \includegraphics[width=0.85\textwidth]{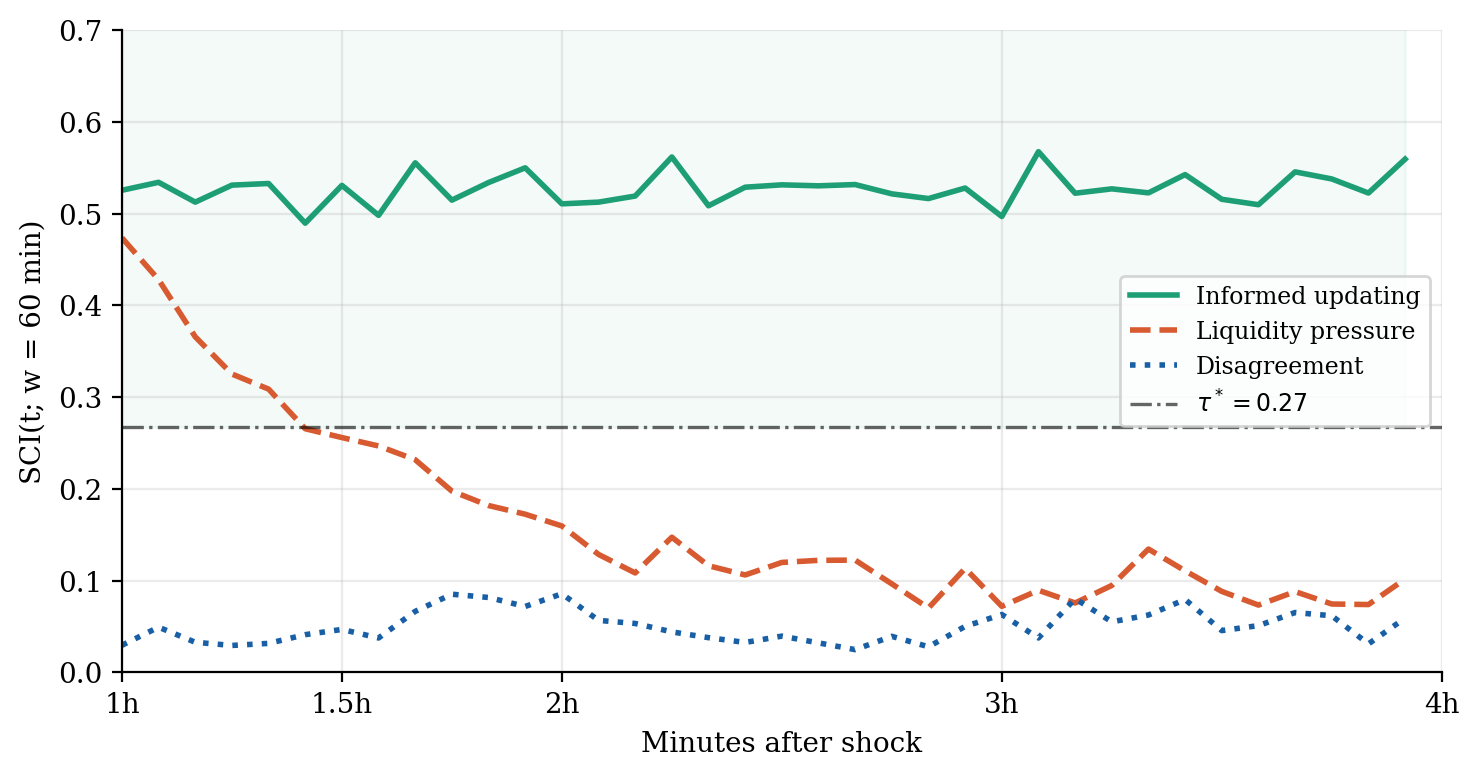}
  \caption{Time-varying $\SCI(t; w = 60\text{ min})$ under three
  representative regimes.
  The persistence ratio $\PR(t,w)$ underlying the index is well-defined
  on the rolling window.}
  \label{fig:tvsci}
\end{figure}

Figure~\ref{fig:tvsci} illustrates the time-varying SCI under three
representative regimes with $w = 60$-minute rolling windows.

\section{Illustrative Application: 2024 Election Shocks}\label{sec:empirical}

We apply the framework to the three major 2024 U.S.\ presidential
election shocks documented in \citet{tsang2026political}, framed
explicitly as illustrative application rather than empirical
recalibration.

The values reported in Table~\ref{tab:shocks} are computed on
\emph{synthetic paths simulated from the DGPs of
Section~\ref{sec:montecarlo}} that match the qualitative microstructure
descriptions in \citet{tsang2026political} for each event:
spike-with-reversal for the debate, persistent post-shock level for the
assassination attempt, and two-sided trading with little net repricing
for the Biden withdrawal.
We are explicit that this is \textbf{not} an empirical application based
on reconstructed on-chain data: the trader-level signed-flow series
required to compute $\HHI^{\text{flow}}$ from blockchain transactions
was not available to us at submission, and the values below are
simulation-derived.
The qualitative ordering of the three events is consistent with the
framework, but should be treated as illustrative until replaced by
ledger-derived estimates.
A subsequent revision will compute these values from the on-chain
ledger reconstruction documented in \citet{tsang2026anatomy}.

\begin{table}[H]
  \centering
  \caption{Illustrative application: full component decomposition for
  three 2024 election shocks}
  \label{tab:shocks}
  \small
  \begin{tabular}{l ccc c c >{\raggedright\arraybackslash}p{3.4cm}}
    \toprule
    \textbf{Event} & $\PR$ & $\TS$ & $\HHI^{\text{flow}}$
       & $\SCI$ & $> \tau^{*}$? & \textbf{Predicted regime} \\
    \midrule
    Debate (Jun 28)        & 0.22 & 0.17 & 0.10 & 0.165 & No  & Liquidity pressure \\
    Assassination (Jul 13) & 0.51 & 0.11 & 0.01 & 0.448 & Yes & Informed updating \\
    Biden dropout (Jul 21) & 0.15 & 0.97 & 0.02 & 0.005 & No  & Disagreement \\
    \bottomrule
  \end{tabular}
\end{table}

The full decomposition makes the framework's reasoning transparent.
The debate shock has moderate but not high $\PR$ (the spike-then-reversal
pattern produces partial reversal) and moderate $\HHI$, but consensual
$\TS$; the composite falls below threshold mainly because $\PR$ is too
low.
The assassination attempt has high $\PR$ (the persistent post-shock
price level), low $\TS$, and very low $\HHI$; all three components
support informed-updating classification.
The Biden dropout has low $\PR$ \emph{and} extremely high $\TS = 0.97$
(fully two-sided trading); even though $\HHI$ is low, the consensus
component $(1 - \TS) = 0.03$ is what suppresses the SCI to near zero.
This is the signature of a politically salient event with a
microstructurally weak trading response --- precisely the situation
the SCI is designed to detect.

\section{Implementation Guide}\label{sec:implementation}

\subsection{Required Data}

Three time-stamped data streams aggregated to 5-minute bins:
(i) Price series $\{p_{t}\}$;
(ii) Aggressive volume $\{B_{t}, S_{t}\}$ classified by initiating side
(\citet{lee1991inferring} tick rule when order books are not observable);
(iii) Trader-level signed flow $\{\Delta v_{j,t}\}$ from on-chain ledger
reconstruction.
For Polymarket, all three streams can be reconstructed from public
Polygon chain transaction data, supplemented where necessary by
order-book or trade-signing conventions for the aggressive-volume
classification; \citet{tsang2026anatomy} document the relevant contract
addresses and decoding procedures.

\subsection{Algorithm}

\begin{algorithm}[H]
  \caption{Compute SCI for a Shock Event}\label{alg:sci}
  \begin{algorithmic}[1]
    \Require Logit-prices $\{\ell_{t}\}$, signed volumes $\{B_{t}, S_{t}\}$,
             trader flows $\{\Delta v_{j}\}$, shock time $t_{s}$, window $W$
    \State $\mathcal{W} = [t_{s}, t_{s} + W]$
    \State \textbf{If} $\sum_{\tau \in \mathcal{W}} |\ell_{\tau} - \ell_{\tau-1}| < \epsilon$
           \textbf{return} $\SCI = 0$ \Comment{No-trade window}
    \State $\PR \gets |\ell_{t_{s}+W} - \ell_{t_{s}}| / \sum_{\tau} |\ell_{\tau} - \ell_{\tau-1}|$
    \State $B \gets \sum_{t \in \mathcal{W}} B_{t}$;
            $S \gets \sum_{t \in \mathcal{W}} S_{t}$
    \State $\TS \gets 1 - |B - S| / (B + S)$
    \State Apply multi-wallet clustering protocol (below) to
           $\{\Delta v_{j}\}$
    \State $\HHI^{\text{flow}} \gets \sum_{j} (|\Delta v_{j}| / \sum_{j'} |\Delta v_{j'}|)^{2}$ on clustered traders
    \State \Return $\SCI \gets \PR \cdot (1 - \TS) \cdot (1 - \HHI^{\text{flow}})$
  \end{algorithmic}
\end{algorithm}

\subsection{Multi-Wallet Clustering Protocol}

The \texttt{coord\_manip\_broad} failure mode (Section~\ref{subsec:exp2})
makes this section critical, not merely a caveat.
We recommend a four-step clustering protocol applied before computing
$\HHI^{\text{flow}}$:

\begin{enumerate}[noitemsep,label=(\alph*)]
  \item \textbf{Common funder identification:} cluster wallets receiving
        their initial USDC deposit from the same source address. This
        catches the most common multi-wallet operator pattern.
  \item \textbf{Temporal co-movement:} cluster wallets whose trading
        activity is highly synchronized (share $> 50\%$ of trading bins
        with identical signed direction over a 30-day pre-shock window).
  \item \textbf{Custodial filter:} exclude or down-weight known exchange
        and custody wallet addresses, which represent aggregated
        pseudo-participants.
  \item \textbf{Graph community detection:} for the residual set, apply
        Louvain or Leiden community detection to the wallet-funding
        graph.
\end{enumerate}

Each step is partial; together they substantially reduce breadth
inflation.
None fully resolves the issue without ground-truth KYC data.
A market-level robustness check is to compute $\HHI^{\text{flow}}$ both
with and without clustering and report the difference: a large gap
indicates that the raw breadth measure is being inflated by
multi-wallet operators.

\subsection{Caveats}

(i) \textbf{$\TS$ vs.\ market-making}: where order-book data are
available, restrict $\TS$ to aggressive-side trades.
(ii) \textbf{Cross-platform aggregation}: compute SCI separately per
platform; differences themselves measure cross-platform credibility
divergence.
(iii) \textbf{Bootstrap inference}: for confidence intervals, resample
at the trade level and recompute $B = 500$ times.

\section{Limitations and Future Work}\label{sec:future}

This revision incorporates substantial methodological improvements
relative to the original specification, but several limitations remain
that point to clear research priorities.

\paragraph{Validation scope.}
The Monte Carlo evidence establishes discrimination among designed
microstructure regimes, not external validation of either coordination
credibility (an unobserved theoretical construct) or downstream
behavioral response (the substantively important quantity).
A high-priority next step is constructing a labeled real-world dataset
of 50--100 prediction-market shocks with independent measures of
downstream coordination response: post-shock changes in donation flow,
media coverage volume, social-media engagement, and institutional
repositioning.
Such a dataset would enable proper external validation and weight
calibration.
\textbf{This is the single most important direction for follow-up work.}

\paragraph{The coordination-vs-information tension.}
The two failure modes identified in Experiment 2 reflect a genuine
tension between two related but non-identical targets: information
content and coordination credibility.
The current SCI specification embeds a strong commitment to the
coordination-credibility target (penalizing all concentration).
Domains where information content is the operative concern (e.g.,
financial-market high-frequency applications where rational arbitrageurs
will respond to informed signals regardless of their concentration)
may benefit from a different weighting --- in particular, a positive
$\HHI$ contribution rather than a negative one.
Future work should formalize the conditions under which each weighting
is appropriate.

\paragraph{Coordinated manipulation and clustering.}
The \texttt{coord\_manip\_broad} failure mode is the most actionable
adversarial pattern for blockchain markets.
It can in principle be addressed through the multi-wallet clustering
protocol of Section~\ref{sec:implementation}, but the protocol's
effectiveness depends on the manipulator's sophistication.
A strong manipulator can route funds through multiple intermediary
addresses to defeat common-funder clustering, can stagger trading
times to defeat temporal co-movement clustering, and can use mixers
to defeat graph community detection.
A combined SCI/forensic-pipeline architecture --- where the SCI
provides initial classification and a separate forensic clustering
analysis flags suspicious wallet networks for manual review --- is
likely the practical path forward.

\paragraph{$\TS$ refinement.}
Two-sidedness as currently specified cannot separate aggressive
disagreement from passive liquidity provision without order-book data.
A refined $\TS^{\text{aggressive}}$ computed only on lift-the-offer
or hit-the-bid trades would be substantively closer to the index's intent.

\paragraph{$\PR$ measures directness, not smoothness or jump timing.}
The persistence ratio captures whether net signed movement dominates
gross absolute movement on the window, but it is invariant to the
\emph{shape} of the path within the window.
A single large jump followed by a flat tail and a smooth gradual
repricing of equal magnitude can yield identical $\PR$ values, even
though their coordination implications may differ (an immediate jump
gives downstream actors less time to react than a gradual move).
The time-varying SCI of Section~\ref{sec:timevarying} partially
addresses this by tracking $\SCI(t;w)$ over short rolling windows
within the post-shock period, but a richer path-shape diagnostic ---
for example, the proportion of total movement occurring in the first
$k$ bins --- would be a useful complement to $\PR$ in future
specifications.

\paragraph{Sequential shocks.}
The framework treats each shock as analytically isolated.
Sequential shocks (a shock occurring during the post-shock window of
an earlier shock) require careful window-overlap handling that we do
not specify here.

\paragraph{Adaptive thresholds.}
The threshold $\tau^{*} = 0.27$ is calibrated on the simulation universe.
A natural extension is a domain-adaptive threshold that updates as
labeled data accumulate, perhaps via a Bayesian online-learning
specification.

\section{Conclusion}\label{sec:conclusion}

This revision addresses the methodological criticisms of the original
specification: the variance-ratio identification problem in short
windows is resolved by the persistence ratio $\PR(t,w)$;
log returns on bounded probabilities are replaced by logit returns;
static position concentration is replaced by post-shock flow
concentration;
the strong ``if and only if'' threshold claim is reformulated as an
operational classifier hypothesis;
and the validation now rests on multiple lines of evidence including a
new adversarial DGP designed to expose coordinated multi-wallet
manipulation.

The validation establishes that the SCI discriminates well among
designed microstructure regimes, with a baseline AUC of $0.984$ on the
canonical three-DGP setup.
On the harder eight-DGP combined set including adversarial regimes,
the SCI achieves $\AUC = 0.847$ --- lower than a flexible logistic
regression benchmark ($0.908$), but superior to additive and
single-component interpretable baselines.
Two failure modes are documented honestly: a Type II conservative bias
on informed-but-concentrated whale activity, and a Type I detection
failure on coordinated multi-wallet manipulation.
The two failure modes are not symmetric and cannot be simultaneously
eliminated within the index's microstructure-only specification.

The SCI's most important contribution is structural rather than
empirical: it provides a transparent, interpretable, and operationally
feasible diagnostic that decomposes the question ``is this price move
credible as a coordination signal?'' into three measurable components
with clear behavioral interpretations.
Whether prediction markets function as coordination devices, as
distortions of public expectations, or as ordinary forecasts is not
answered by a single index.
But identifying which of these regimes a specific shock belongs to is
exactly what the SCI is designed to do, and the validation evidence
shows it does this well within its defined scope.

\section*{Generative AI Disclosure}
\addcontentsline{toc}{section}{Generative AI Disclosure}

In preparing this manuscript, the author used Anthropic's Claude
Opus 4.7 for copy-editing and for rendering figures from numerical
data.
All methodology, analysis, and conclusions are the author's own;
the author reviewed and edited all AI-generated content and takes
full responsibility for the final manuscript.

\bibliographystyle{apalike}

\appendix

\section{DGP Specifications}\label{app:dgp-spec}

All DGPs simulate four hours of 5-minute bins ($n_{\text{bins}} = 48$),
starting from $p_{0}^{+} = 0.72$ post-shock.
$\mathrm{Gamma}(k, \theta)$ uses the shape-scale convention with mean
$k\theta$.
Random seed: \texttt{20260429}.

\begin{table}[H]
  \centering
  \caption{Full DGP specifications}
  \label{tab:dgp-spec}
  \scriptsize
  \begin{tabular}{l p{3.4cm} p{2.7cm} p{2.4cm} l}
    \toprule
    \textbf{DGP} & \textbf{Logit return process} & \textbf{Buy/sell volumes}
       & \textbf{Traders, Dir.\ $\alpha$} & \textbf{Label} \\
    \midrule
    \texttt{informed}        & $r_{t} \sim \mathcal{N}(0.0010, 0.0022^{2})$, iid
       & B: $\Gamma(2.5, 5\!\times\!10^{4})$, S: $\Gamma(0.5, 1.5\!\times\!10^{4})$
       & $N \in [150,250]$, $\alpha = 4$ & 1 \\
    \addlinespace
    \texttt{liquidity}       & $r_{t} = -0.55\,r_{t-1} + \mathcal{N}(-0.0008, 0.0025^{2})$
       & B: $\Gamma(0.6, 1.5\!\times\!10^{4})$, S: $\Gamma(2.5, 4\!\times\!10^{4})$
       & $N \in [20,50]$, $\alpha = 0.4$ & 0 \\
    \addlinespace
    \texttt{disagreement}    & $r_{t} \sim \mathcal{N}(0, 0.0040^{2})$
       & total: $\Gamma(2,4\!\times\!10^{4})$, split $\sim$Beta(8,8)
       & $N \in [60,120]$, $\alpha = 1.2$ & 0 \\
    \addlinespace
    \texttt{whale\_informed} & as informed
       & as informed
       & $N \in [4,8]$, $\alpha = 0.3$ & 1 \\
    \addlinespace
    \texttt{noisy\_broad}    & $r_{t} \sim \mathcal{N}(0, 0.006^{2})$
       & total: $\Gamma(2,4\!\times\!10^{4})$, split $\sim$Beta(3,3)
       & $N \in [120,200]$, $\alpha = 3$ & 0 \\
    \addlinespace
    \texttt{manip\_then\_info} & $t<6$: $r_{t} = -0.5 r_{t-1} + \mathcal{N}(-0.0005, 0.003^{2})$; $t\geq 6$: $r_{t} \sim \mathcal{N}(0.0012, 0.0022^{2})$
       & switches at $t=6$
       & $N \in [80,150]$, $\alpha = 2.5$ & 1 \\
    \addlinespace
    \texttt{persistent\_two\_sided} & $r_{t} \sim \mathcal{N}(0.0008, 0.002^{2})$
       & total: $\Gamma(2.5, 4\!\times\!10^{4})$, split $\sim$Beta(8,8)
       & $N \in [80,150]$, $\alpha = 2$ & 0 \\
    \addlinespace
    \texttt{coord\_manip\_broad} & $r_{t} \sim \mathcal{N}(0.0009, 0.0024^{2})$, iid
       & B: $\Gamma(2.5, 5\!\times\!10^{4})$, S: $\Gamma(0.4, 1.2\!\times\!10^{4})$
       & $N \in [80,130]$, $\alpha = 4$ & 0 \\
    \bottomrule
  \end{tabular}
\end{table}

\end{document}